\documentclass[twocolumn,aps,prl,showpacs,floatfix,preprintnumbers,amsmath,amssymb,superscriptaddress,10pt]{revtex4-1}

\usepackage{amsmath}
\usepackage{amssymb}
\usepackage{graphicx}
\usepackage{dcolumn} 
\usepackage{bm} 
\usepackage{undertilde}
\usepackage{color}
\usepackage{natbib}

\newcommand{\mean}[1]{\left < #1 \right >}
\newcommand{\abs}[1]{\left | #1 \right |}
\newcommand{\fourier}[2]{{\hat{#1}}_{#2}}
\newcommand{\nabcp}{\utilde{\nabla}}

\begin{document}
\title{{Vortex arrays and meso-scale turbulence of self-propelled particles}}

\author{Robert Gro{\ss}mann}
\author{Pawel Romanczuk}
\author{Markus B{\"a}r}
\affiliation{Physikalisch-Technische Bundesanstalt Berlin, Abbestr. 2-12, 10587 Berlin, Germany}

\author{Lutz Schimansky-Geier}
\affiliation{Department of Physics, Humboldt-Universit{\"at} zu Berlin, Newtonstr. 15, 12489 Berlin, Germany}

\begin{abstract}
Inspired by the Turing mechanism for pattern formation, we propose a simple self-propelled particle model with short-ranged alignment and anti-alignment at larger distances. It is able to produce orientationally ordered states, periodic vortex patterns as well as meso-scale turbulence. The latter phase resembles observations in dense bacterial suspensions. The model allows a systematic derivation and analysis of a kinetic theory as well as hydrodynamic equations for density and momentum fields. A phase diagram with regions of such pattern formation as well as spatially homogeneous orientational order and disorder is obtained from a linear stability analysis of these continuum equations. Microscopic Langevin simulations of the self-propelled particle system are in agreement with these findings. 
\end{abstract}

\date{\today}
\maketitle

The term active matter refers to non-equilibrium systems of interacting, self-propelled entities which are able to take up energy from their environment and convert it into motion \cite{romanczuk_active_2012,marchetti_hydrodynamics_2013}. Examples, such as cytoskeletal filaments \cite{julicher_active_2007}, chemically driven colloids \cite{theurkauff_dynamic_2012} or flocks of birds \cite{ballerini_interaction_2008} have recently received a lot of attention in physics, chemistry and biology. They exhibit a wide range of collective phenomena which are absent in systems at thermodynamic equilibrium, for example large-scale travelling bands and polar clusters \cite{schaller_polar_2010,menzel_traveling_2013} as well as arrays of vortices \cite{riedel_self-organized_2005,sumino_large-scale_2012}. In this context, bacteria represent important model systems, which have been used to investigate such different aspects as clustering \cite{peruani_collective_2012} and rheological properties \cite{sokolov_enhanced_2009} of active matter systems.

Recently, irregular vortex structures were experimentally observed in dense bacterial suspensions \cite{dombrowski_self-concentration_2004,sokolov_enhanced_2009,zhang_swarming_2009,wensink_meso-scale_2012,liu_multifractal_2012}. In addition, a phenomenological model was proposed which describes the observed behavior including the power spectrum of the bacterial dynamics \cite{wensink_meso-scale_2012,dunkel_fluid_2013}. The spectrum at large wave numbers as well as the dynamic vortex patterns in these experiments and simulations are reminiscent of a turbulent state which led the authors to denominate this new phenomenon as ``meso-scale turbulence''. 

We aim to formulate a model at the level of individual particles which is capable to produce such mesoscopic spatiotemporal patterns. Inspired by the Turing mechanism of short-range activation and long-range inhibition in reaction-diffusion systems \cite{turing_chemical_1952,kondo_reaction-diffusion_2010}, we propose interacting self-propelled particles with local alignment at short length scales and anti-alignment at larger distances. Such a type of interactions may be realized by a competition of local alignment and large-scale hydrodynamic back-flow effects \cite{sokolov_physical_2012,tsang_flagella-induced_2014}, e. g. in suspensions of bacterial microswimmers, which leads to preferential alignment of neighboring cells and anti-alignment with more distant swimmers.

Here, we abstain from considering a detailed model of individual swimmers immersed and interacting through a surrounding fluid. Our model includes effective interactions of self-propelled particles of the type first formulated by Vicsek et al. \cite{vicsek_novel_1995}. The original Vicsek model displays surprisingly complex spatiotemporal behavior \cite{chate_collective_2008,romanczuk_mean-field_2012,caussin_emergent_2014}, but it does not exhibit meso-scale turbulence or vortex patterns. In contrast, our generalized model including anti-alignment exhibits both orientationally ordered states and periodic vortex arrays. Between both states, a transition region with spatio-temporal irregular behavior is found. The simulated patterns therein reproduce most features found in the turbulent state reported earlier in experiments and phenomenological macroscopic models \cite{wensink_meso-scale_2012,dunkel_fluid_2013}.

The simplicity of our microscopic model makes it analytically tractable and allows to derive and analyse kinetic equations as well as approximated equations for the coarse-grained density and momentum fields. The validity of the kinetic equations is demonstrated by comparison of their phase diagram to direct simulations of the ``microscopic'' self-propelled particle model. The resulting hydrodynamic model is a modification of the Toner-Tu equations \cite{toner_flocks_1998,toner_reanalysis_2012} similar to the phenomenological model of bacterial turbulence suggested earlier \cite{wensink_meso-scale_2012,dunkel_minimal_2013}. We show explicitly that a non-local anti-alignment interaction is necessary to obtain a negative effective viscosity of the order-parameter field. 

We consider $N$ self-propelled particles moving with constant speed $v_0$ in a two-dimensional system with periodic boundary conditions. Each particle interacts with all neighbors within a finite interaction range $l_{\rm{int}}$. For simplicity, we work in natural units such that the particle mass, speed and interaction range equal one ($m=1$, $v_0=1$, $l_{\rm int}=1$). 

The stochastic equations of motion for the individual particles read  
 \begin{align}
 \label{eqn:microscopic:model}
   \dot { \bf r}_i  = {\bf v}_i,\quad \dot\varphi_i = \sum_{j\neq i}^N {T}_\varphi({\bf r}_{ji},\varphi_i,\varphi_j) + \sqrt{2 D_{\varphi}} \, \zeta_i(t) 
  \end{align}
with ${\bf r}_i$ and ${\bf v}_i=(\cos \varphi_i, \sin \varphi_i)^T$ denoting the position and velocity vector of the $i$-th particle, respectively. Due to the constant speed $\abs{{\bf v}_i}=1$, the velocity vector ${\bf v}_i$ is fully determined by the polar angle $\varphi_i$ representing the direction of motion of the particle. The particles reorient according to pair interactions ${T}_\varphi({\bf r}_{ji},\varphi_i,\varphi_j)$, which depend on the distance vector ${\bf r}_{ji} = {\bf r}_j - {\bf r}_i$ and the orientation angles of the interaction partners. The last term on the right-hand side of the angle equation \eqref{eqn:microscopic:model} represents angular noise with intensity $D_\varphi$. Here, $\zeta_i(t)$ describes Gaussian random processes with zero mean and $\mean{ \zeta_i(t) \zeta_j(t')} = \delta_{ij} \, \delta (t - t')$. 

\begin{figure}
 \begin{center}
    \centering
   \includegraphics[width=0.8\columnwidth]{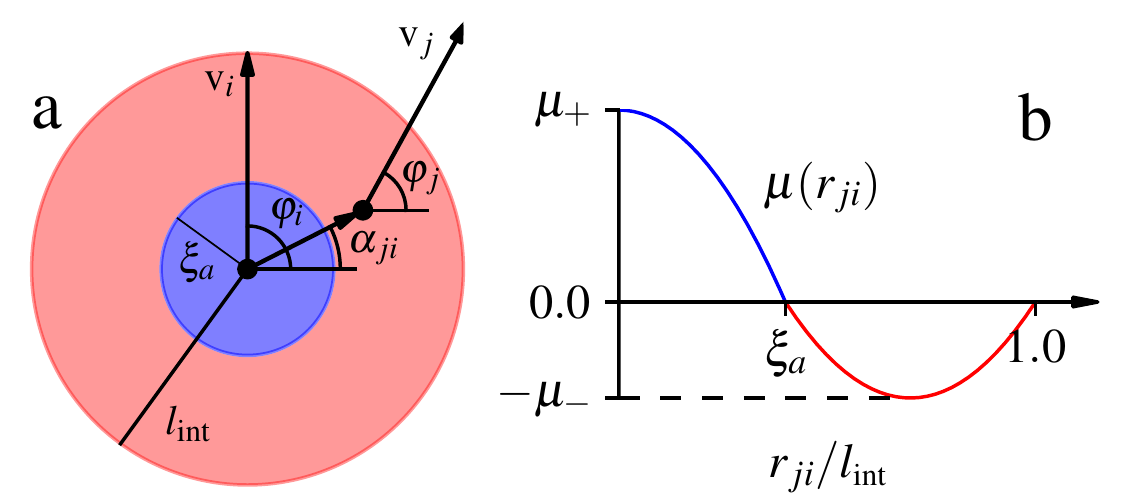}\\
   \includegraphics[width=\columnwidth]{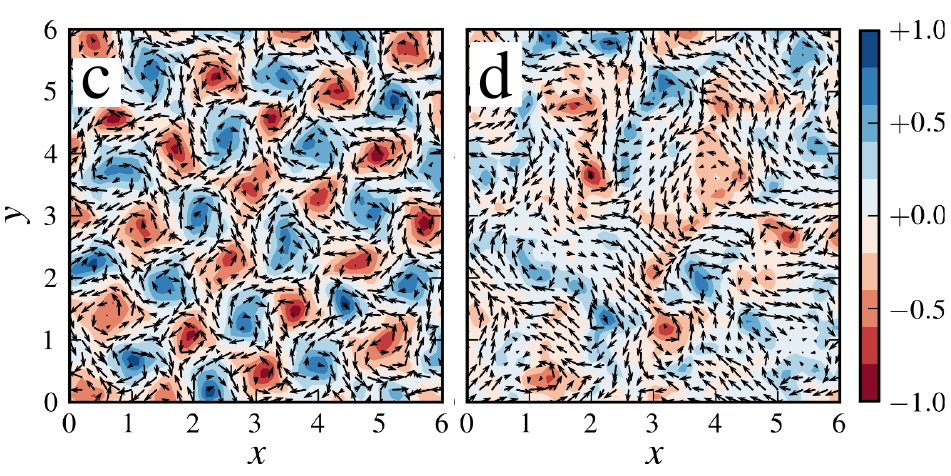}

\caption{\label{fig1}(color online) (a) Interaction scheme in the frame of the focal particle. (b) The alignment strength $\mu(r_{ji})$ versus distance \eqref{eqn:SRA:LRAA}. 
Figs. (c)-(d) show snapshots of the coarse-grained velocity field ${\bf v}$ (coarse graining length $\delta l=\xi_a=0.2$) obtained from numerical solution of the microscopic model \eqref{eqn:microscopic:model}: 
(c) spatially periodic vortex state ($\mu_+=3.2$, $\mu_{-}=3.2\cdot 10^{-2}$) and 
(d) turbulent state close to the transition from the vortex phase to onset of polar order ($\mu_+=1.6$, $\mu_{-}=8 \cdot 10^{-3}$). 
The color bar indicates the local vorticity $\nabla \wedge {\bf v} = \partial_x v_y - \partial_y v_x$ of the velocity field ${\bf v}$. Simulation parameters: $L=30$, $N=135000$, $D_\varphi=1.0$, $\xi_a=0.2$, $\xi_r = 0.1$, $\kappa = 10$, $\Delta t = 10^{-2}$.
}
 \end{center}
\end{figure}

The pair-wise interaction is modelled as follows:
 \begin{align}
 \label{eq:torque}
	 {T}_\varphi \! \left( {\bf r}_{ji},\varphi_i,\varphi_j \right ) =  & + \mu \! \left(r_{ji} \right ) \sin \left ( \varphi_j - \varphi_i \right) \\ &  - \kappa \sin \left ( \alpha_{ji} - \varphi_i \right ) \Theta (\xi_r - r_{ji}). \nonumber 
 \end{align}
The first term aligns the interacting particles either parallel or anti-parallel depending on the sign of the interaction strength $\mu(r_{ji})$, which is a function of the scalar distance $r_{ji}=|{\bf r}_{ji}|$. For $\mu(r_{ji})>0$, the velocity vectors align, whereas $\mu(r_{ji})<0$ implies an anti-parallel orientation of these vectors (anti-alignment). The distance dependence of the alignment interactions is depicted in Fig. \ref{fig1}a-b. For simplicity, we model $\mu(r_{ji})$ as a piecewise parabolic function 
\begin{align}
  \label{eqn:SRA:LRAA}
  \mu(r_{ji}) & = \begin{cases} + \mu_{+}   \left(1 - \left( r_{ji}/{\xi_a} \right)^{2} \right) &  \mbox{for} \;\;\,  0 \le r_{ji} \le \xi_a, \\
				     - \mu_{-}  \frac{4\left (r_{ji} - \xi_a\right )\left(1-r_{ji}\right)}{\left(1-\xi_a\right)^2}  &  \mbox{for} \;\;\,  \xi_a < r_{ji} \le 1
		 \end{cases}
 \end{align}
with $\mu_\pm>0$. Thus, the interaction favors alignment at short distances and anti-alignment at larger distances within the interaction range (Fig. \ref{fig1}a). The maximal strengths of alignment and anti-alignment are denoted by $\mu_+$ and $-\mu_-$, respectively (Fig. \ref{fig1}b). 

The second term on the right-hand side of Eq. \eqref{eq:torque} describes a repulsive soft-core interaction where $\alpha_{ji}=\rm{arg}({\bf r}_{ji})$ is the polar positional angle of particle $j$ in the reference frame of particle $i$. A constant repulsion strength $\kappa\geq0$ is assumed below a distance $r_{ji}\le\xi_r$ and no interaction for $r_{ji}>\xi_r$, indicated by the Heaviside function $\Theta(x)$. The repulsive interaction is motivated by steric interaction of finite-sized particles. We assume that short-ranged alignment and repulsion act on comparable length scale $\xi_r\lesssim\xi_a$ and use $\xi_r=\xi_a/2$.

\begin{figure}
 \begin{center}
 \includegraphics[width=0.99\columnwidth]{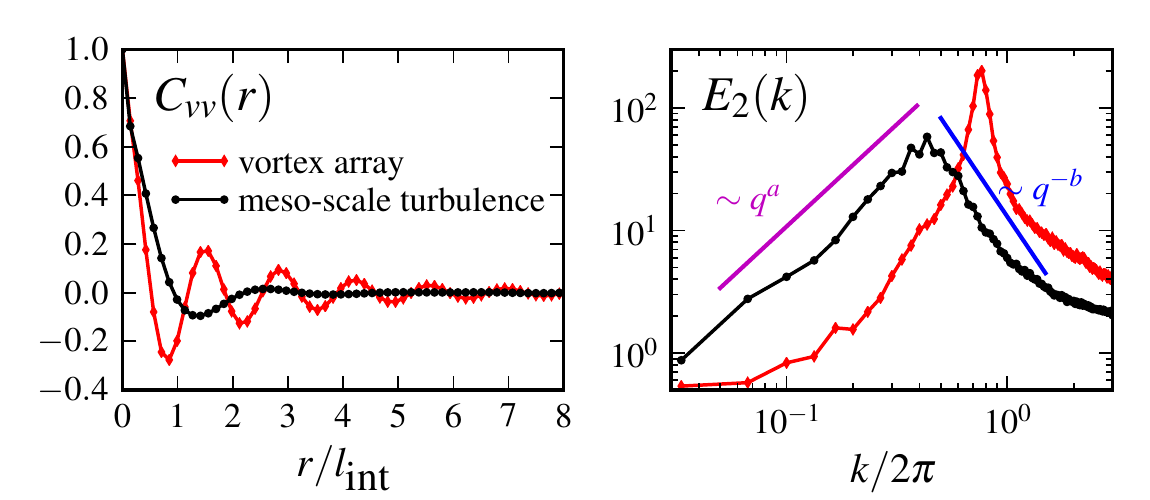}
 \caption{(color online) Two-point velocity correlation $C_{vv}(r)$ and its Fourier transform $E_2(k)$, cf. \cite{wensink_meso-scale_2012} for definitions, for a periodic array of vortices deep in region III ($\mu_+=3.2$, $\mu_{-}=3.2\cdot 10^{-2}$, red) and in the ``turbulent'' state close to the critical line between region II and III ($\mu_+=1.6$, $\mu_{-}=8\cdot 10^{-3}$, black). The solid lines in the right panel indicate the scalings observed in dense {\em Bacillus subtilis} suspensions ($a \approx 1.7$, $b \approx 2.7$) and the corresponding phenomenological theory \cite{wensink_meso-scale_2012}. The shape of the energy spectrum is not universal but depends on the choice of model parameters. Simulation parameters: $L = 30$, $N=135000$, $D_\varphi = 1.0$, $\xi_a = 0.2$, $\xi_r = 0.1$, $\kappa = 10$, $\Delta t = 10^{-2}$. \label{fig2}}
 \end{center}
\end{figure}

\begin{figure}
 \begin{center}
 \includegraphics[width=\columnwidth]{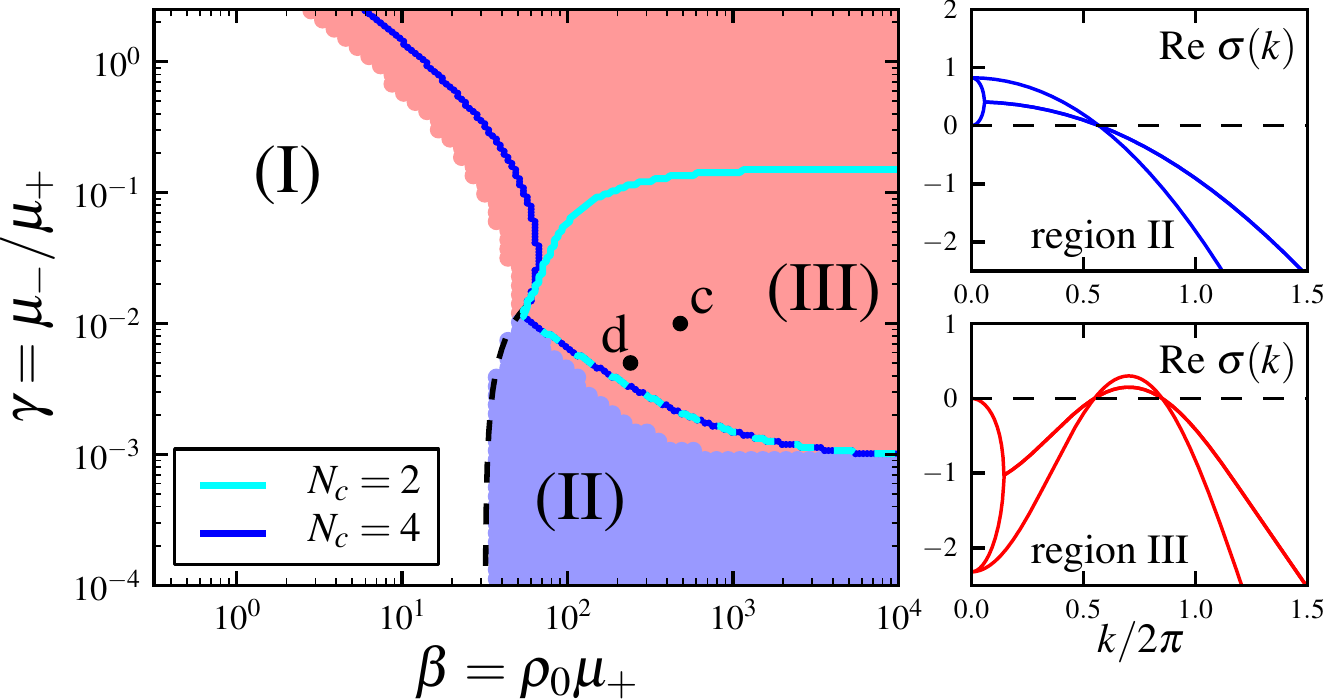}
 \caption{(color online) {\bf Left:} Phase diagram as predicted by the kinetic theory: disordered state (I, white), polar order  (II, blue) and periodic patterns (III, red). The solid lines are critical lines between region (I) and (III) obtained from hydrodynamic theory for different truncations $N_c$ of the interaction operator \eqref{eqn:def:hat:mu:D}. Colored dashed lines indicate the corresponding boundary of region (II) and (III) as predicted by the hydrodynamic theory. The black dashed line represents the boundary $\text{Re}(\sigma(k=0))=0 \, \Leftrightarrow \, \pi \rho_0 \mu_0 - D_\varphi =0$, independent of $N_c$. The labelled points indicate the parameter values corresponding to the snapshots shown in Figs. \ref{fig1}c-d. Parameters: $\rho_0=150$, $D_{\varphi}=1$, $\xi_a = 0.2$, $\xi_r = 0.1$, $\kappa = 10$. 
{\bf Right:} Examples of dispersion relations $\text{Re}(\sigma(k))$ from hydrodynamic theory for region II (top, $\beta=60$, $\gamma=10^{-3}$), and region III (bottom, $\beta=70$, $\gamma=5\cdot10^{-2}$); Other parameters as in the phase diagram.  
\label{fig3}}
 \end{center}
\end{figure}

We performed numerical simulations of the microscopic particle dynamics \eqref{eqn:microscopic:model} in the high density regime $\rho_0 \gg 1$. Examples of the observed patterns are shown in Fig. \ref{fig1}c,d and movies can be found in Supplementary Material (SM). For low $\mu_+$ and $\mu_-$, we observe a spatially homogeneous, disordered state (region I in the phase diagram Fig. \ref{fig3}, Movie 1 in SM). For low $\mu_-$, an increase in $\mu_+$ eventually leads to the onset of long-ranged orientational order (region II in Fig. \ref{fig3}, movie 2 in SM). In contrast, for large $\mu_+$ and large $\mu_-$ we observe periodic arrays of vortices in the velocity field (Fig. \ref{fig1}c, region III in Fig. \ref{fig3}, Movie 3 in SM). These vortices are typically accompanied by a periodic modulation of the density. However, a strong short-range repulsion (large $\kappa$) enforces a quasi-homogeneous density profile with only weak fluctuations.  The pattern corresponds to the one observed in the phenomenological model by Dunkel et al. \cite{dunkel_minimal_2013}. Interestingly, for parameters in the vortex phase (III) close to the emergence of polar order (II), the short-ranged alignment induces mesoscopic convective flows, whereas $\mu_-$ is just sufficiently large to break global orientational order. These flows destroy the spatial periodicity of the vortex array, and the emergent, irregular pattern is best described as \textit{meso-scale turbulence} \cite{wensink_meso-scale_2012} (Fig. \ref{fig1}d, Movie 4 in SM). Fig. \ref{fig2} shows the two-point velocity correlation $C_{vv}(r)$ and its Fourier transform $E_2(k)$, see \cite{wensink_meso-scale_2012} for definitions, for a periodic vortex array and in the turbulent phase. In the former case, we typically observe damped oscillations of $C_{vv}(r)$ corresponding to a sharp peak in the energy spectrum $E_2(k)$ indicating spatially periodic vortex patterns. In the turbulent phase, $C_{vv}$ shows only a single minimum. The qualitative behavior of the velocity correlation function in the turbulent regime is in good agreement with observations made in dense bacterial suspensions \cite{dombrowski_self-concentration_2004,sokolov_enhanced_2009,zhang_swarming_2009,wensink_meso-scale_2012,liu_multifractal_2012}. 

Following \cite{farrell_pattern_2012,grossmann_self-propelled_2013}, we derive the dynamics of the one-particle density function $p({\bf r},\varphi,t) = \mean{\sum_{i=1}^N \delta({\bf{r}} - {\bf{r}}_i(t)) \delta(\varphi - \varphi_i(t))}$ starting from the Fokker-Planck equation \cite{risken_fokker-planck_1996} of the $N$-particle probability density function (PDF). The evolution of $p({\bf r},\varphi,t)$ depends on the two-particle distribution function due to binary interactions. Here, we approximate the two particle PDF by neglecting correlations between particles. Using this choice of closure, we obtain a nonlinear Fokker-Planck equation for the one-particle density: 
 \begin{align}
  \label{eqn:nonl:FPE}
  & \partial_t p \!\left ( {\bf r},\varphi,t \right ) =  - {\bf v} \cdot \nabla p \!\left ( {\bf r},\varphi,t\right ) + D_\varphi \partial^2_\varphi p \!\left ( {\bf r},\varphi,t\right )  \\
									&- \partial_\varphi \left[ \iint \text{d}^2 r' \text{d}\varphi' \, {T}_\varphi \! \left( {\bf r}',\varphi,\varphi' \right ) p \!\left ({\bf r},\varphi,t \right ) p \!\left ({\bf r}+{\bf r}',\varphi',t \right ) \right] \! .\nonumber 
 \end{align} 
 
Eq. \eqref{eqn:nonl:FPE} is solved by the spatially homogeneous, isotropic state $p_0 = \rho_0/(2 \pi)$ where $\rho_0$ is the spatially homogeneous particle density. In order to study the stability of this solution, it is convenient to work in Fourier space with respect to the angular variable $\varphi$. 
We derive equations of motion for the Fourier coefficients $\fourier{f}{n}({\bf r},t) = \int_0^{2 \pi}  \text{d}\varphi \, e^{i n \varphi} \, p({\bf r},\varphi,t)$ from \eqref{eqn:nonl:FPE}. Further, we expand $p \! \left ({\bf r}+{\bf r}',\varphi',t \right )$ into a Taylor series around ${\bf r}'=\mathbf{0}$, which allows for the transformation of the interaction integral in Eq. \eqref{eqn:nonl:FPE} into an infinite series of differential operators \cite{belintsev_pattern_1981,hernandez_clustering_2004}. The Fourier transformation of Eq. \eqref{eqn:nonl:FPE} then yields
 \begin{align}
 \label{eqn:dyn:FT:angle}
 \partial_t \fourier{f}{n} = & - \left ( \nabcp \fourier{f}{n-1} + \nabcp^* \fourier{f}{n+1} \right)- n^2 D_{\varphi} \fourier{f}{n} \\ 
					      & + n \pi \! \left [ \fourier{f}{n-1} \, \hat{\mu}_{\Delta} \, \fourier{f}{1} - \fourier{f}{n+1} \, \hat{\mu}_{\Delta} \, \fourier{f}{-1} \right ] \nonumber \\
     & - n \pi  \left [ \fourier{f}{n-1} \, \mathcal{K}_{\Delta} \nabcp \fourier{f}{0} - \fourier{f}{n+1} \, \mathcal{K}_{\Delta} \nabcp^{*} \fourier{f}{0}  \right ] \nonumber
 \end{align}
where $\nabcp = \left ( \partial_x + i \partial_y \right)/2$.  

Some terms in Eq. \eqref{eqn:dyn:FT:angle} are also present for non-interacting particles: the first two terms on the right hand side in parenthesis account for the convection due to active motion, whereas the third term describes the diffusion of the direction of motion due to angular fluctuations. The term containing the differential operator $\hat{\mu}_{\Delta}= \int_0^{\infty}  \text{d}r \, r \mu(r) J_0 \! \left ( r \sqrt{-\Delta} \,\right )$ originates in the alignment interaction. It can be rewritten using the series expansion of the Bessel function of the first kind $J_0(x)$ to 
 \begin{align}
 \label{eqn:def:hat:mu:D}
 \hat{\mu}_{\Delta} & = \sum_{n=0}^{\infty} \mu_n \Delta^{\!n} \, , \quad \mu_n = \int_0^{\infty}  \text{d}r \, \mu(r) \, \frac{r^{2n+1}}{4^n (n!)^2}\, .
 \end{align}
The last two terms stem from the repulsion interaction in Eq. \eqref{eq:torque}. The operator $\mathcal{K}_{\Delta}$ is given by the series 
 \begin{eqnarray}
	 \label{eqn:def:K:D}
	 \mathcal{K}_{\Delta} = \kappa \sum_{n=0}^\infty \frac{\xi_r^{2n+3}}{4^{n} n! (n+1)! (2n+3)} \, \Delta^{\!n} .   
 \end{eqnarray}

Starting from Eq. \eqref{eqn:dyn:FT:angle}, we analyzed the linear stability of the disordered, homogeneous state $p_0$. The number of parameters determining the linear stability of $p_0$ is reduced by introducing $\beta = \rho_0 \mu_+$, $\gamma = {\mu_-}/{\mu_+}$ and $\eta = \rho_0 \kappa$. The results of the linear stability analysis (phase diagram) are shown in Fig. \ref{fig3}. The different behavior of the largest eigenvalues $\sigma(k)$, where $k=|{\bf k}|$ denotes the wavenumber, reveals the existence of three distinct phases. We observe an instability towards a spatially homogeneous orientionally ordered state indicated by a maximum of $\text{Re}(\sigma(k))$ at $k=0$. For other parameter values, we find a novel instability with a maximum of $\text{Re}(\sigma(k))$ at a finite wavenumber $k\neq0$ predicting the emergence of a spatial pattern with a characteristic length scale. The destabilization of $p_0$ at finite $k$ is only possible if $\gamma$ is large enough. This particular instability cannot be found in a system with alignment only, because it crucially depends on the presence of sufficiently strong anti-alignment interaction. 

We derive approximated equations for the relevant observables: density $\rho({\bf r},t)$ and the momentum field ${\bf w}({\bf r},t)$, related to $\fourier{f}{0}$ and $\fourier{f}{1}$ \cite{Note1}. The dynamics of the $n$-th Fourier coefficient $\fourier{f}{n}$ is in general coupled to the coefficients $\fourier{f}{n-1}$ and $\fourier{f}{n+1}$, cf. Eq. \eqref{eqn:dyn:FT:angle}. Thus, the nonlinear, non-local Fokker-Planck equation \eqref{eqn:nonl:FPE} is equivalent to an infinite hierarchy of dynamical equations \eqref{eqn:dyn:FT:angle} in Fourier space. This formulation allows for a systematic investigation of the consequences of the following two approximations: (i) elimination of Fourier coefficients $\fourier{f}{n}$ with $\abs{n} > n_c> 0$, i.e. assuming fast relaxation of high order orientational modes; (ii) elimination of higher-order derivatives in \eqref{eqn:def:hat:mu:D} and \eqref{eqn:def:K:D} by focusing only on large-scale spatial dynamics.  

First, we eliminate the second Fourier mode by assuming $\partial_t \fourier{f}{2} \approx 0$ \cite{bertin_hydrodynamic_2009,grossmann_self-propelled_2013} but keep the full operator $\fourier{\mu}{\Delta}$ accounting for the non-local interactions. For simplicity, only the corresponding linearized evolution equations are considered (linearized hydrodynamic equations)
\begin{subequations} 
\label{eqn:hydro:theor:real}
 \begin{align}
	 \frac{\partial \rho}{\partial t}   = &  - \nabla \cdot {\bf w} , \\
   \frac{\partial {\bf w}}{\partial t} = & - c {\nabla \rho} - D_\varphi {\bf w} + \pi \rho_0 \hat{\mu}_{\Delta} {\bf w} + \frac{\Delta {\bf w}}{16 D_\varphi} \, ,
\label{eqn:hydro:theor:real:w}
 \end{align}
\end{subequations}
where the positive coefficient $c = \left ( 1/2 + \pi \eta \xi_r^3/6 \right )$ is proportional to the compressibility. 

A comparison of the prediction on linear stability from the full kinetic theory in Eq. \eqref{eqn:dyn:FT:angle} with corresponding results from Eqs. \eqref{eqn:hydro:theor:real} shows that the latter theory qualitatively reproduces the general structure of the phase space (Fig. \ref{fig3}). Excellent quantitative agreement is obtained for large noise strengths due to the strong damping of higher order Fourier modes and an effectively weak coupling of different modes (not shown). 

So far, we have kept the full interaction operator $\fourier{\mu}{\Delta}$ defined in Eq. \eqref{eqn:def:hat:mu:D}. We can now connect our theory to phenomenological theories as the one by Toner and Tu \cite{toner_flocks_1998,toner_reanalysis_2012} or the one by Dunkel et al. \cite{wensink_meso-scale_2012,dunkel_minimal_2013} by truncating the infinite series in Eq. \eqref{eqn:def:hat:mu:D} as $\hat{\mu}_{\Delta,N_c} = \sum_{n=0}^{N_c} \, \mu_n \Delta^n$. The stability of the linearized dynamics at short length scales (large $k$) requires $(-1)^{N_c} \mu_{N_c} < 0$ to hold. If the anti-alignment interaction is dominant, this implies even $N_c$ and the lowest order truncation is given by $N_c=2$. In this case, the linearized hydrodynamic equation of the momentum field reads 
\begin{align}
	\label{eq:w:Nc2}
    \frac{\partial {\bf w}}{\partial t} \approx & - c{\nabla \rho} + \left [ \pi \rho_0 \mu_0 -D_\varphi \right ] \! {\bf w}  \\&+  \underbrace{\left [ \pi \rho_0 \mu_1 + (16D_\varphi)^{-1} \right ]}_{=\Gamma} \! \Delta {\bf w} +  \pi \rho_0\mu_2 \Delta^{\!2} {\bf w} \, , \nonumber
 \end{align}
where $\Gamma$ denotes the effective viscosity of the momentum field $\mathbf{w}$. One sees from Eq. \eqref{eq:w:Nc2}, that polar order increases locally, if $\pi \rho_0 \mu_0 -D_\varphi > 0$ (right of black dashed line in the phase diagram, Fig. \ref{fig3}). 

We performed the linear stability analysis of the spatially homogeneous, isotropic state for different truncation orders $N_c$. The results are depicted in Fig. \ref{fig3}. Qualitatively, the three phases of the system -- disordered, polar order and vortex arrays -- are predicted by the truncated, linearized hydrodynamic theory for all $N_c \ge 2$. The predictions of the linear stability analysis by the hydrodynamic theory on the transition lines are, however, only for $N_c \ge 4$ in rough quantitative agreement with the predictions of the full kinetic theory \eqref{eqn:dyn:FT:angle}. In contrast, the predicted structure of the phase space for $N_c = 2$ differs significantly from the one obtained from the full kinetic theory: the extent of the vortex phase is much smaller than predicted by the kinetic theory. 

However, it is possible to gain important qualitative insights in the mechanism leading to periodic structures in the order-parameter field, which is related to a change of sign of the effective viscosity $\Gamma$ in Eq. \eqref{eq:w:Nc2}, as suggested already in \cite{dunkel_minimal_2013}. In turn, this is only possible if the alignment strength is negative, i.e. anti-alignment is present. 

Since the presented model captures the phenomenology of meso-scale turbulence (cf. Fig. \ref{fig1}d and Fig. \ref{fig2}), we hypothesize that this state may emerge due to a competition of short-ranged alignment and anti-alignment at larger distances suppressing the emergence of long-ranged, spatially homogeneous orientational order. 
Moreover, the introduced model exhibits regular periodic vortex array patterns if the anti-alignment interaction is strong enough.

In summary, we have proposed a simple model of self-propelled particles with purely local interactions: alignment of close-by neighbors and anti-alignment with particles at larger distances, which exhibits not only a polar ordered phase but also periodic vortex arrays and meso-scale turbulence. The latter emerges close to the critical line for the onset of polar order, where increasing convective flows destroy the regular vortex pattern. We were able to establish a direct connection between the microscopic and mesoscopic behavior of interacting self-propelled particles by a derivation and analysis of coarse-grained equations suggested to describe these novel phases of active matter. On a more general level, the emergence of collective motion patterns in self-propelled particle systems due to the simultaneous action of short-range alignment and anti-alignment at larger distances represents the analogue of the Turing mechanism based on short-range activation and long-range inhibition in reaction-diffusion systems.  

We thank S. Heidenreich for enlightening  discussions and gratefully acknowledge the support by the DFG via GRK 1558 and IRTG 1740. PR acknowledges the hospitality of the Kavli Institute for Theoretical Physics (UCSB) and support in part by the National Science Foundation under Grant No. NSF PHY11-25915..  

\bibliographystyle{Myapsrev}	
\bibliography{SPAA}

\end{document}